\documentclass[a4paper,twoside]{article}
\newif\ifcheck
\checktrue
\newcommand{\affil}[1]{$^{\rm #1}$}
\usepackage[authoryear]{natbib}
\bibpunct{(}{)}{;}{a}{}{,}
\usepackage{graphicx}

\usepackage{lscape}
\pdfoutput=1
\usepackage{epstopdf}
\usepackage{amsmath}
\usepackage{amssymb}
\usepackage{rotating}
\usepackage{appendix}
\usepackage{color}

\usepackage{tabularx}
\usepackage{tabulary}

\date{} 
\newcommand{\kms}{\mbox{km\,s$^{-1}$}}
\newcommand{\about}{$\sim\!\!$~}
\newcommand\ion[2]{#1$\;${%
\ifx\@currsize\normalsize\small \else
\ifx\@currsize\small\footnotesize \else
\ifx\@currsize\footnotesize\scriptsize \else
\ifx\@currsize\scriptsize\tiny \else
\ifx\@currsize\large\normalsize \else
\ifx\@currsize\Large\large
\fi\fi\fi\fi\fi\fi
\rmfamily\@Roman{#2}}\relax}%
\def\arcdeg{\hbox{$^\circ$}}

\def\arcsec{\hbox{$^{\prime\prime}$}}

\def\magarcsec2{\ \rm{mag\ arcsec}^{-2}}
\def\micron{\hbox{$\mu$m}}
\def\reff@jnl#1{{\rm#1\/}}
\def\apj{\reff@jnl{ApJ}}
\def\aj{\reff@jnl{AJ}}                  
\def\araa{\reff@jnl{ARA\&A}}            
\def\apj{\reff@jnl{ApJ}}                
\def\apjl{\reff@jnl{ApJ}}               
\def\apjs{\reff@jnl{ApJS}}              
\def\aap{\reff@jnl{A\&A}}               
\def\mnras{\reff@jnl{MNRAS}}            
\def\prd{\reff@jnl{Phys.Rev.D}}         
\def\prl{\reff@jnl{Phys.Rev.Lett}}      
\def\pasp{\reff@jnl{PASP}}              
\def\nat{\reff@jnl{Nature}}             
\def\iaucirc{\reff@jnl{IAU~Circ.}}%
\def\aaps{\reff@jnl{A\&AS}}%
\def\apss{\reff@jnl{Ap\&SS}}%
\newcommand\actaa{\ref@jnl{Acta Astron.}}%
\newcommand\pasa{\ref@jnl{PASA}}%
\newcommand\sovast{\ref@jnl{Soviet~Ast.}}%

\def\Fp{F}
\def\dm15{\ifmmode{\Delta m_{15}}\else{$\Delta m_{15}$}\fi}
\newcommand{\Msun}{\>{\rm M_{\odot}}}

\title{\large\bf\flushleft Light Echoes of Transients and Variables 
in the Local Universe}
\author{\parbox{\textwidth}{\flushleft
\vspace{-0.5cm}
{\it A. Rest\affil{A,C}, B. Sinnott\affil{B}, \& D. L. Welch\affil{B}}\\
\vspace{0.4cm}
{\small \affil{A}\,Space Telescope Science Institute, 3700 San Martin
Dr., Baltimore, MD 21218, USA}\\
{\small \affil{B}\,Department of Physics and Astronomy, McMaster University,
Hamilton, Ontario L8S 4M1, Canada}\\
{\small \affil{C}\,Email: contact arest@stsci.edu}
}}
\begin{document}
\twocolumn[
\begin{changemargin}{.8cm}{.5cm}
\begin{minipage}{.9\textwidth}
\vspace{-1cm}
\maketitle
%
%
\small{\bf Abstract:}
Astronomical light echoes, the time-dependent light scattered by dust
in the vicinity of varying objects, have been recognized for over a
century. Initially, their utility was thought to be confined to
mapping out the three-dimensional distribution of interstellar
dust. Recently, the discovery of spectroscopically-useful light echoes
around centuries-old supernovae in the Milky Way and the Large
Magellanic Cloud has opened up new scientific opportunities to exploit
light echoes.

In this review, we describe the history of light echoes in the local
Universe and cover the many new developments in both the observation
of light echoes and the interpretation of the light scattered from
them. Among other benefits, we highlight our new ability to
spectroscopically classify outbursting objects, to view them from
multiple perspectives, to obtain a spectroscopic time series of the
outburst, and to establish accurate distances to the source event.  We
also describe the broader range of variable objects whose properties
may be better understood from light echo observations. Finally, we
discuss the prospects of new light echo techniques not yet realized in
practice.


\medskip
\medskip
\end{minipage}
\end{changemargin}
]
\small

\section{Introduction}

Modern astrophysics has long benefited from the additional information
conveyed by variable stars. Some of the most physically interesting
sources are those which emit pulses of light during eruptive or
disruptive events which traditionally have resulted in brief temporal
windows during which critical physical information about rare,
transitional phases can be acquired. In recent years, the ability to
extend those temporal windows through detecting and analyzing light
echoes from historical events has been realized and a number of
eruptions without contemporaneous photometry and spectroscopy have
become observable with modern instrumentation.

``Light echoes'' (hereafter, LEs) are simply faint reflections of the
light from a bright astrophysical source off material - in these cases,
circumstellar or interstellar dust. The sizes and irregular surfaces
of typical dust grains results in incident light being scattered over
a wide range of angles although forward-scattering is typically far
stronger than back-scattering.  The scattered light received by an
observer on Earth has taken a longer path than any light which may
have been received directly from the outbursting source, and so it
arrives later. This time-delay depends on the geometry between the
dust location and the source event (Section~\ref{sec:geometry}). As we
shall describe presently, the extreme brightness of supernovae (SNe)
have allowed detection of their LEs many centuries after the observed
event. In the intervening time, we have developed physics and our
instrumentation capabilities to a point where we can gain tremendous
new insights on the nature of these rare events - a situation which
would be impossible without the temporal delay introduced by LEs.

This review reports and summarizes the wealth of new information which
has become available as a result of studying LEs from varying
astrophysical sources, focusing on scattered light echoes in the local
Universe. It is now over a century since the first LEs were discovered
around Nova Persei 1901 \citep{Ritchey01b,Ritchey01a,Ritchey02}, and
recognized as such by \citet{Kapteyn02,Perrine03}. Initially, LEs were
mainly used to constrain the 3D-position of the scattering dust and
its properties like grain-size distribution, density, and
composition. The impact of unresolved light echoes on the spectra and
light curves of SNe was recognized as a complication in the analysis
of extra-galactic SNe.  We now recognize that currently
observationally-difficult or impossible problems of asymmetry
estimation, spectral typing, lightcurve reconstruction, and more
precise distances yield to LE observations.  From the wealth of
applications listed above, we emphasize in this review the more recent
techniques which use LEs as a means to directly observe the outburst
light of an event by using a non-direct line-of-sight to the
source. The full potential for the use of LE techniques has not yet
been realized and we identify key exploration opportunities for the
near future.

\section{Finding Light Echoes}

The first LEs were discovered around Nova Persei 1901
\citep{Ritchey01b,Ritchey01a,Ritchey02}, and it was interpreted as
such by \cite{Kapteyn02,Perrine03}. Since then, LEs have been seen
associated with a wide variety of objects: the Galactic Nova
Sagittarii 1936 \citep{Swope40}, the eruptive variable V838
Monocerotis \citep{Bond03}, the Cepheid RS Puppis
\citep{Westerlund61,Havlen72}, the T Tauri star S~CrA \citep{Ortiz10},
and the Herbig Ae/Be star R~CrA \citep{Ortiz10}. Echoes have also been
observed from extragalactic SNe, with SN~1987A being the most famous
case \citep{Crotts88,Suntzeff88}, but also including SNe~1980K
\citep{Sugerman12}, 1991T \citep{Schmidt94,Sparks99}, 1993J
\citep{Sugerman02,Liu03}, 1995E \citep{Quinn06}, 1998bu
\citep{Garnavich01,Cappellaro01}, 2002hh \citep{Welch07,Otsuka12},
2003gd \citep{Sugerman05,VanDyk06,Otsuka12}, 2004et \citep{Otsuka12},
2006X \citep{Wang08,Crotts08}, 2006bc \citep{Gallagher11,Otsuka12},
2006gy \citep{Miller10}, and 2007it \citep{Andrews11}.  One factor
that all these objects have in common was that the LEs were detected
serendipitously in the presence of the still luminous stars or
transients.  These light echoes may influence the spectra as well as
the light curves of the sources when unresolved, in particular for
events like type II SNe that are likely located in dust-rich
environments \citep[e.g.,][]{Schaefer87b,Chugai92,DiCarlo02,Otsuka12},
but have also been observed for SN~199T \citep{Schmidt94}, a SN~Ia
that is not necessarily expected to be in such a dust-rich environment.
\citet{Roscherr00} finds that light echoes cannot explain the
extremely long decline of SN~IIn like SN~1988Z and SN~1997ab.

The suggestion that historical SNe might be studied by their scattered
LEs was first made by \cite{Zwicky40}.  Simple scaling arguments
\citep{Shklovskii64,vandenBergh65a,vandenBergh66} based on the
visibility of Nova Persei predicted LEs from SNe as old as a few
hundred to a thousand years can be detected, especially if the
illuminated dust has regions of high density ($\geq 10^{-8}
\rm{cm}^{-3}$). 

The few dedicated surveys in the last century for LEs from historic
SNe \citep{vandenBergh65b,vandenBergh65a,vandenBergh66,Boffi99} and
novae \citep{vandenBergh77,Schaefer88} have been
unsuccessful. However, these surveys did not use digital image
subtraction techniques \citep{Tomaney96,Alard98,Alard00} to remove the
dense stellar and galactic backgrounds. Even the bright echoes near
SN~1987A \citep{Crotts88,Suntzeff88} at $V\approx 21.3\magarcsec2$ are
hard to detect relative to the dense stellar background of the Large
Magellanic Cloud (LMC). \citet{Maslov00} suggested that using a
wide-field polarization imager to detect LEs of historic SNe.

The situation changed with the advent of CCDs and telescopes with
large field-of-views, which enabled the astronomical community to
entertain the first wide-field time-domain surveys with sufficient
depth. The first LEs of ancient events were found serendipitously by
\cite{Rest05b} as part of the SuperMACHO survey \citep{Rest05a}: They
found LEs associated with 400--900 year-old SNRs.  This result
demonstrated that the LEs of historical SNe and other transients could
be found, and subsequent targeted searches found LEs of Tycho's SN
\citep{Rest07,Rest08b}, Cas~A \citep{Rest07,Rest08b,Krause08a}, and
$\eta$~Carinae \citep{Rest12_etac}.  A recent search for LEs
of historic SNe around 4 recent SNe in M83 using polarization imaging
was not successful \citep{Romaniello05}.  It should be noted that
\citet{Krause05} discovered ``IR echoes'' from the Cas A SNR, in which
the dust absorbs the outburst light, is warmed, and re-radiates light
at longer infrared (IR) wavelengths.  This is different to the usual
LE phenomenon described here, where the light is simply scattered by
dust, therefore preserving the spectral characteristics of the source
event.

LEs are extended objects, often with very faint surface
brightnesses. Therefore it is necessary to apply difference imaging
\citep{Tomaney96,Alard98,Alard00} to separate them from the sky
background in order to be able to identify them and measure their
surface brightness. However, only the relative fluxes between two
epochs is revealed in a difference image. Therefore obtaining absolute
fluxes for individual epochs has traditionally relied on a single
template image that is free of LEs, which often must be constructed by
a complicated and usually subjective process of hand-selecting
suitable images \citep[e.g.,][]{Sugerman05b}.  \cite{Newman06}
presented a solution to this issue by applying the NN2 method of
\cite{Barris05} to extract the relative fluxes of LEs across a range
of epochs directly from a series of difference images. This method
treats all images the same and makes maximal use of the observational
data. The efficacy of this technique was demonstrated by applying it
to the LEs around SN~1987A \citep{Newman06}.  The most common source
of false LE candidates is scattered/reflected light from bright stars
falling in the focal plane beyond the edges of the detectors. Small
pointing errors between image epochs can produce difference features
whose shapes and surface brightnesses can mimic LEs. These fairly
common optically-induced difference features make it difficult to
identify faint LEs with software and most LE searches rely at least in
part on visual inspection of the difference images. With the advent of
the next generation of wide-field, time domain surveys like Pan-STARRS
\citep{Kaiser10}, PTF \citep{Rau09}, Skymapper \citep{Keller07}, and
ultimately LSST \citep{Ivezic08}, visual inspection will no longer be
a viable option and software solutions for identifying true light
echoes will need to be developed.

\section{Light Echo Primer}

There are many excellent reviews of LEs covering the various aspects
of LE science. The groundwork of LE geometry was laid by
\cite{Couderc39}, and the surface brightness of LEs was derived by
\citet{Chevalier86}. After SN~1987A started to show its beautiful set
of LEs, more derivations of the surface brightness of single-scattered
LEs \citep[e.g.,][]{Schaefer87a,Xu94,Sugerman03,Patat05} and
multiple-scattered LEs \citep{Patat05,Patat06} were done, as well as
its impact on observed SN spectra and light curves discussed
\citep[e.g.,][]{Schaefer87b,Patat05,Patat06}.  \citet{Sugerman03}
investigated the range of circumstances which would produced
observable light echoes for different classes of transients.

Besides the scattered light echoes which preserve the SED of the
source event, there are other types of time-delay which are often
referred to as echoes.  One of them is the so-called ``IR echo'',
where dust absorbs the outburst light, is warmed, and re-radiates
light at longer infrared (IR) wavelengths.  IR echoes have been
observed around SNe \citep[e.g.,][]{Bode80a,Dwek83,Krause05,Kotak09}
and novae \citep[e.g.,][]{Bode80b,Bode85,Gehrz88}.  ``Recombination''
echoes are another form of echoes where the initial light gets
absorbed and then re-radiated at a different wavelength
\citep[e.g.,][]{Panagia91,Gould94,Gould98}.  The ionization light echo
of a quasar in ``Hanny's Voorwerp'' was discovered by the Galaxy Zoo
project \citep{Lintott09,Rampadarath10}.  All of the above are
different to the usual LE phenomenon described here, where the light
is simply scattered by dust, therefore preserving the spectral
characteristics of the source event. Because IR and recombination
echoes only share the geometry with scattered light echoes, we will
focus on scattered light echoes, and only refer to the other echoes
when they are relevant to a particular object with scattered light
echoes.

\subsection{Geometry\label{sec:geometry}}
\begin{figure}[t]
\includegraphics[width=210pt]{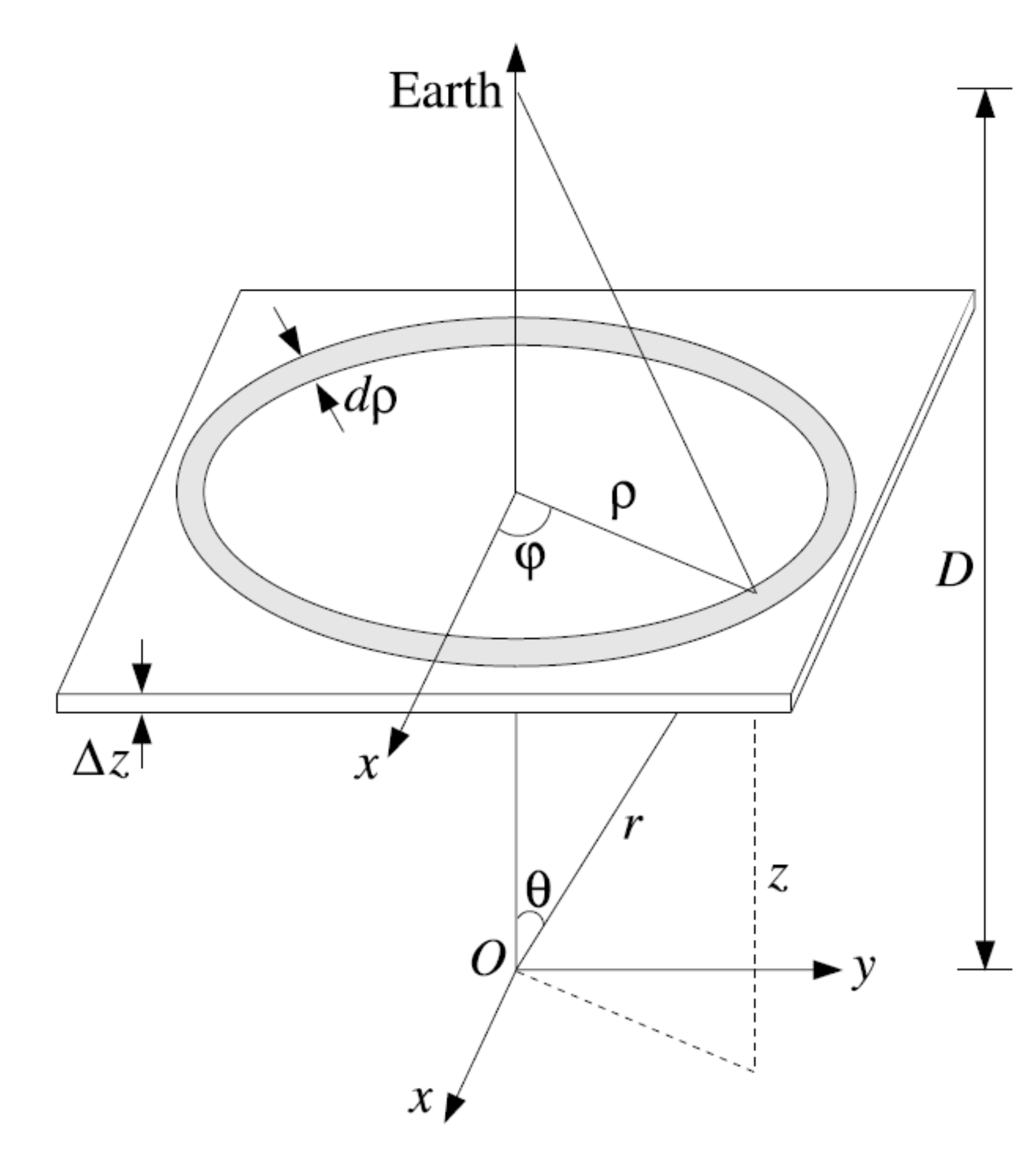}%
\caption[]{Figure from \cite{Sugerman03}, which was adapted from
\cite{Xu94}. Note that distances are not to scale and $D$ is much
larger than $z$.
\label{fig:geometry}}
\end{figure}

Figure~\ref{fig:geometry} (from \cite{Sugerman03}, adapted from
\cite{Xu94}) illustrates the geometry of the LE
phenomenon. Light originating from the SN event source, $O$, is
scattered by dust located a distance $z$ in front of the SN and a
projected distance $\rho$ perpendicular to the line of sight, and is
redirected towards the observer. The event source and observer are
separated by a distance $D$. In application
$z<<D$, and the LE equation \citep{Couderc39}
\begin{equation}
z = \frac{\rho^2}{2ct} - \frac{ct}{2} \label{eq:le}
\end{equation}
can be derived, where $t$ is the time since the explosion was
originally observed and $c$ is the speed of light. Then the distance
$r$ from the scattering dust to the source event is $r^2=\rho^2+z^2$,
and the projected distance on the sky is $\rho = (D-z) \tan\gamma$,
where $\gamma$ is the angular separation between the source event and
the scattering dust.

Measuring $\rho$ through imaging, we can determine the exact
three-dimensional location of the scattering dust by simply knowing
the time since the outburst occurred and the distance to the
object. In the case that the distance to the object is not well known,
the above geometry still provides relative distances between distinct
scattering dust locations. Through LE imaging alone, the
three-dimensional dust structure in front of an outburst event can be
mapped in great detail. While the dust structure can be measured by
observing multiple scattering dust locations, it is most thoroughly
mapped through monitoring the apparent motion of a given LE system,
since the apparent motion is heavily dependent on the scattering dust
inclination (see Section~\ref{sec:appmot}).
\subsection{Surface Brightness}

The surface brightness of LEs has been derived in different variants
by various authors
\citep[e.g.,][]{Chevalier86,Schaefer87a,Xu94,Sugerman03,Patat05}. Here
we follow a derivation of the surface brightness flux $f_{SB}$ by
\cite{Sugerman03} and define
\begin{eqnarray}
f_{SB}(\lambda,t,\phi) = \Fp(\lambda)n(\textbf{r})\left( \frac{c\Delta z}{4\pi r
\rho\Delta\rho} \right)S(\lambda,\theta).
\label{eq:surfacebrightness}
\end{eqnarray}
This equation has four principal components: (1) $\Fp(\lambda)$ is the
lightcurve-weighted integrated event flux of the varying source event,
(2) $n(\textbf{r})$ is the dust density as a function of position
$\textbf{r}$, (3) a wavelength-independent spatial component, where
$\Delta \rho$ and $\Delta z$ are the width of the LE on the sky and
the depth of the scattering dust sheet along the line of sight,
respectively, and (4) the wavelength-dependent integrated scattering
function, $S(\lambda,\theta)$, which we describe in more detail in
Section~\ref{sec:S}.

Note that the event flux $\Fp(\lambda)$ is the flux {\it difference}
of the source with respect to its quiescent state. In all the cases
where the source event is a one-time transient (e.g. SNe) or where the
event is several magnitudes brighter than the magnitude in the
quiescent state (e.g. V 838 Mon), this differentiation is not
important since the quiescent flux is either zero or close to
zero. However, for other objects like Cepheids this is important since
what we can detect is the appearing and disappearing scattered flux
which is due to the brightness variation (i.e. the difference in flux
between its maximum and minimum brightness) and not the total
brightness. For example, we can find scattered light from a
non-varying source at a given position, but it is constant with time
(i.e. there is no apparent motion of the scattered light) and is
commonly called a reflection nebula.

It is important to note that this derivation assumes a photon is only
scattered once (single-scattering approximation). However, for very
dense dust environments multiple scattering can become important
\citep{Chevalier86,Patat05}. The effect that unresolved single and
even more so multiple scattered LEs have on observed colors and
spectra of SNe is discussed in detail in \citet{Patat06}.


\subsubsection{Dust Scattering\label{sec:S}}

The integrated scattering function $S(\lambda,\theta)$ describes the
wavelength-dependent effect of the scattering off dust grains, and is
therefore of great importance when calculating the surface brightness
of LEs. Here we describe how $S$ can be calculated for our Galaxy and
the Magellanic Clouds.

Since different types of dust grains have different integrated
scattering functions, they must be added together to get the total
integrated scattering function
\begin{equation}
S(\lambda,\theta) = \sum_X S_X(\lambda,\theta),
\end{equation}
where $X$ denotes the dust grain type. Here, $X$ can represent silicon
dust grains, carbonaceous dust grains with a neutral Polycyclic
Aromatic Hydrocarbon (PAH) component, or carbonaceous dust grains with
an ionized PAH component \citep{WD01}.

The integrated scattering function $S_X$ for a dust grain of type $X$
is
\begin{eqnarray}
S_X(\lambda,\theta) = \int Q_{SC,X}(\lambda,a)\sigma_g\Phi_X(\theta,\lambda,a)f_X(a)da
\label{eq:scattering}
\end{eqnarray}
where $Q_X(\lambda,a)$ is the dust grain scattering efficiency and
$f_X(a)$ is the dust grain density distribution.  The radius of the
individual dust grains is given by $a$, with a grain cross-section
$\sigma_g=\pi a^2$.  The Henyey-Greenstein
phase function, $\Phi_X$, is given below, with $g_X(\lambda,a)$ denoting
the degree of forward scattering for a given grain~\citep{HG41}.
\begin{eqnarray}
\Phi_X(\lambda,\theta) = \frac{1-g_X^2(\lambda,a)}{(1+g_X^2(\lambda,a)-2g_X(\lambda,a)\cos{\theta})^{3/2}}
\label{eq:henyey}
\end{eqnarray}
Values for $Q_{X}$ and $g_X(\lambda,a)$ can be derived using tables
provided by
B. T. Draine\footnote{http://www.astro.princeton.edu/$\sim$draine/dust/dust.diel.html}
\citep{Draine84,Laor93,WD01,Li01}.

We use the {\it ``MWG''}, {\it ``LMC avg''}, and {\it ``SMC bar''}
models defined by \cite{WD01}, adopting values for their model
parameter $b_C$ of $5.6\times10^{-5}$, $2\times10^{-5}$, and $0$,
respectively \citep[Table 1 and 3,][]{WD01}. The models consist of a
mixture of carbonaceous grains and amorphous silicate grains.
The dust grain size distribution is
\begin{equation}
f(a) \equiv \frac{1}{n_H}  \frac{dn_{gr}}{da},
\end{equation}
where $n_{gr}(a)$ is the number density of grains smaller than size
$a$ and $n_H$ is the number density of H nuclei in both atoms and
molecules. Note that small carbonaceous grains ($a \leq 10^{-3}
\micron$) are PAH-like while large carbonaceous grains ($a>10^{-3}
\micron$) are graphite-like \citep{Li01}. The size distributions for
carbonaceous dust ($f_{ci}(a)=C_{ion} f(a)$ and $f_{cn}(a)=(1-C_{ion})
f(a))$ can then be calculated using equations 2, 4, and 6 of
\cite{WD01}. Assuming PAH/graphitic grains to be 50\% neutral and 50\%
ionized \citep{Li01}, $C_{ion}=0.5$. Equations 5 and 6 are used for
amorphouse silicate dust, $f_{s}$.
\begin{figure}[t]
\includegraphics[width=220pt]{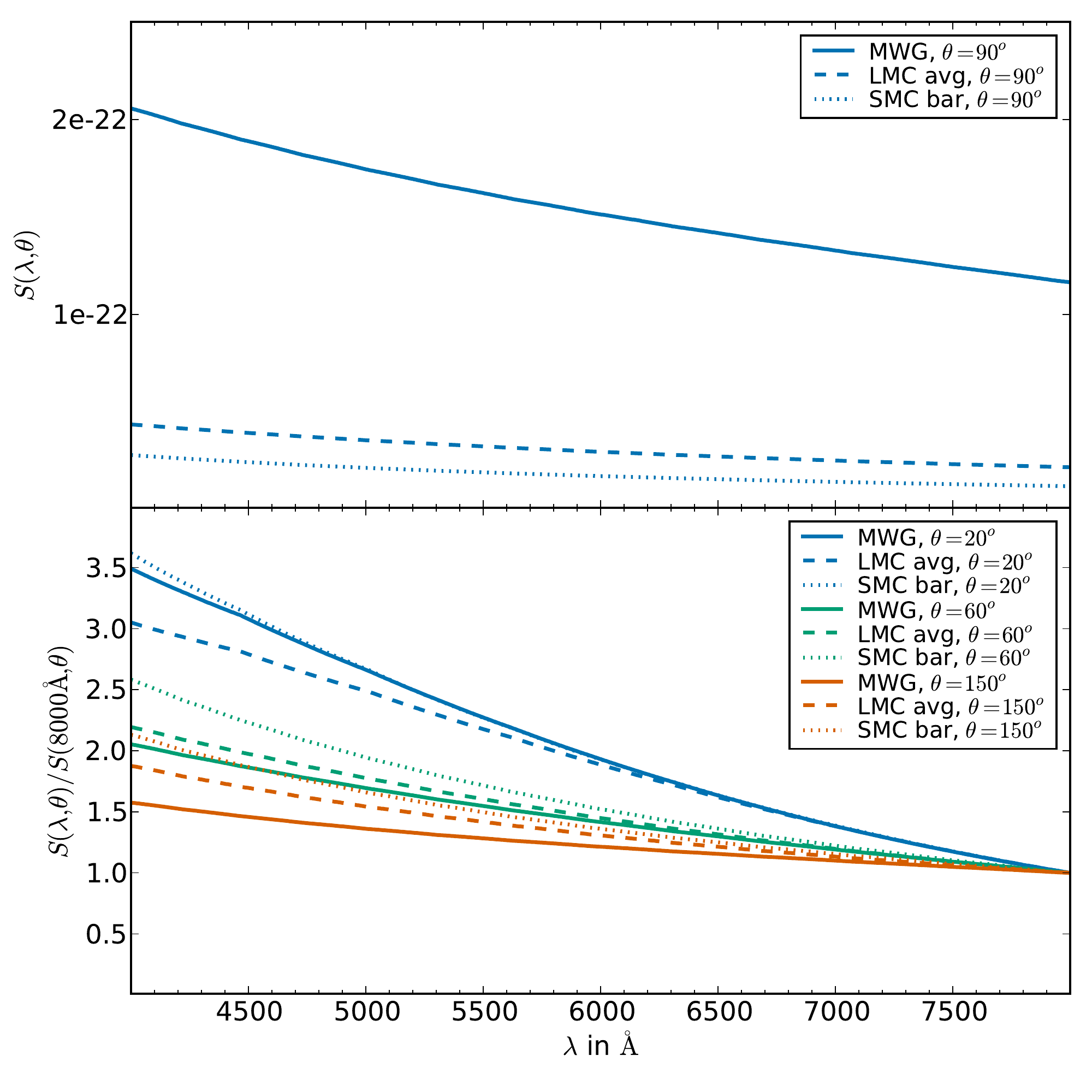}%
\caption[]{The upper panel shows the integrated scattering function
$S(\lambda,\theta)$ for the three dust models {\it ``MWG''}, {\it
``LMC avg''}, and {\it ``SMC bar''}, which are defined in
Section~\ref{sec:S}. The bottom panel shows $S(\lambda,\theta)$
normalized by $S(8000$\AA,$\theta)$ for the three models and various
scattering angles.
\label{fig:S_vs_lambda}}
\end{figure}

\begin{figure}[t]
\includegraphics[width=220pt]{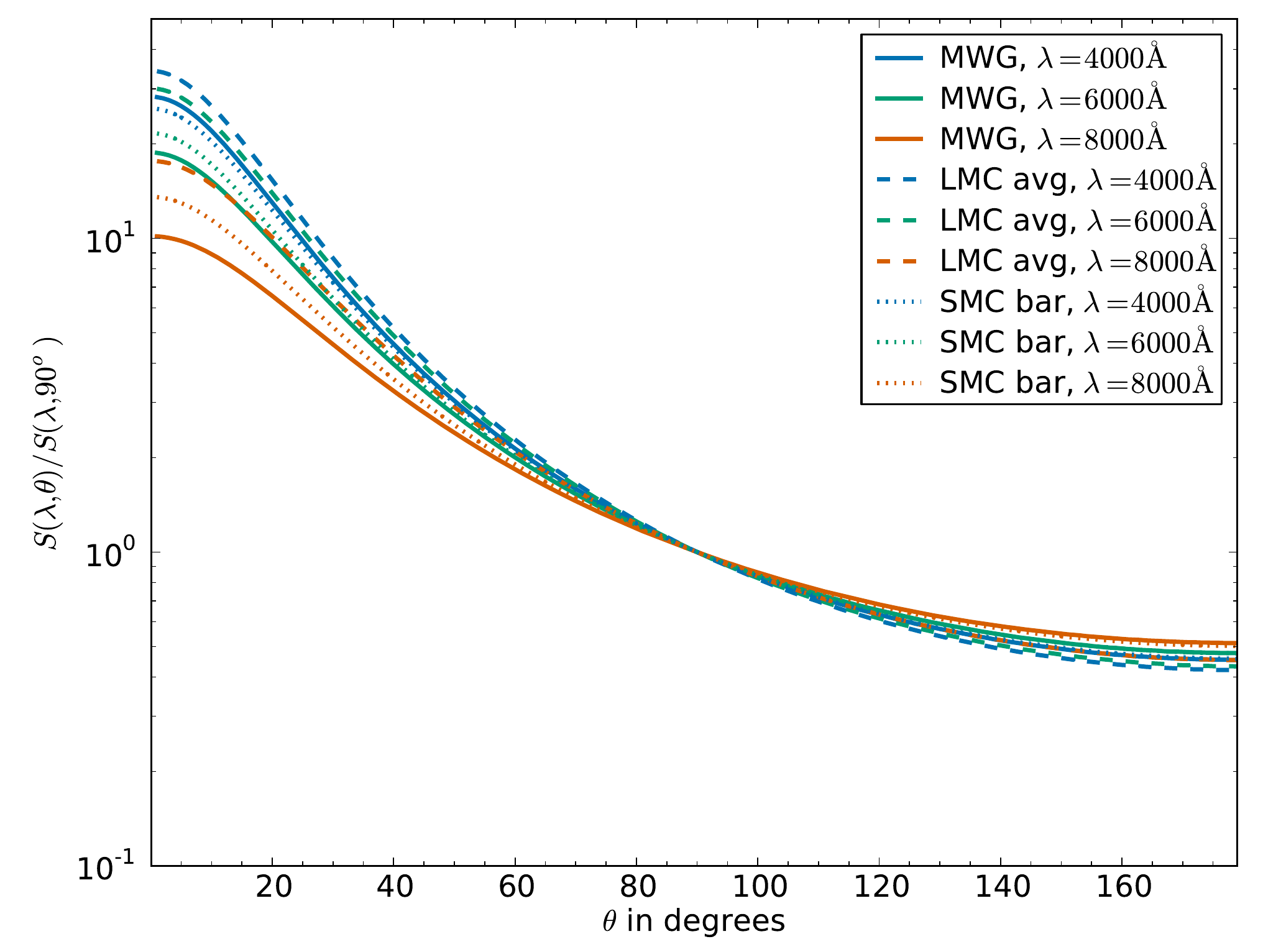}%
\caption[]{$S(\lambda,\theta)$ normalized by $S(\lambda,90^o)$ for the
three dust models {\it ``MWG''}, {\it ``LMC avg''}, and {\it ``SMC
bar''}, which are defined in Section~\ref{sec:S}. Large differences in models
and wavelengths are only apparent for scattering angles $\theta\lesssim60^o$.
\label{fig:S_vs_theta}}
\end{figure}

In the upper panel of Figure~\ref{fig:S_vs_lambda} we show the
integrated scattering function $S(\lambda,\theta)$ as a function of
wavelength for the Galaxy and Magellanic Cloud models, adopting a
scattering angle $\theta=90^{o}$. In the $4000$\AA~to $8000$\AA~region
shown, the Milky Way Galaxy dust scattering is almost twice as
efficient in the blue.  For a detailed comparison of the scattering
efficiencies, we show in the bottom panel of
Figure~\ref{fig:S_vs_lambda} $S(\lambda,\theta)$ normalized by
$S(8000$\AA,$\theta)$ for the three models and various scattering
angles. Note that the difference in blueward scattering efficiency is
much larger between $\theta=20^o$ and $60^o$ than between $60^o$ and
$150^o$. The differences between the models are small, but
significant, and do not seem to vary significantly with scattering
angle. In Figure~\ref{fig:S_vs_theta} we show $S(\lambda,\theta)$ in
log-scale normalized by $S(\lambda,90^o)$ for the different models and
wavelengths. For scattering angles $\theta \gtrsim 60^o$, the
scattering efficiency is quite similar for all models and
wavelengths. For $\theta\lesssim60^o$, however, the efficiency of
blueward scattering is more prominent, with large differences between
different models and different wavelengths clearly visible.

\subsection{Apparent motion}
\label{sec:appmot}

One of the most easily measurable properties of LEs is their
(often superluminal) apparent proper
motion on the sky. The main influences on the apparent motion are the
angular separation between the LE and source event, the time since the
first light of the source reached earth, and the distance to the
source event ($\theta$, $t$, and $D$, respectively). This opens up the
possibility that if either one of the age or distance to the event is
known, the other parameter can be estimated since the angular
separation can be measured very accurately. However, the apparent
motion also depends on the inclination of the scattering dust filament
with respect to the line of sight. This fact is often overlooked or
ignored, and therefore the derivation of age/distance of the event
assumes implicitly or explicitly a certain dust inclination. This
can lead to an underestimate of the systematic uncertainty and
subsequently to a wrong scientific conclusion. A tale of caution is
the analysis by \cite{Krause05} of the IR echoes of Cas A.  Their main
scientific conclusion is that most if not all of these IR echoes are
caused by a recent X-ray flare of the Cas A SNR based on their
apparent motions. However, the analysis was flawed because it did not
account for the fact that the apparent motion strongly depends on the
inclination of the scattering dust filament (\citealt{Dwek08}; Rest
et~al., in prep.).

\begin{figure}[t]
\includegraphics[width=230pt]{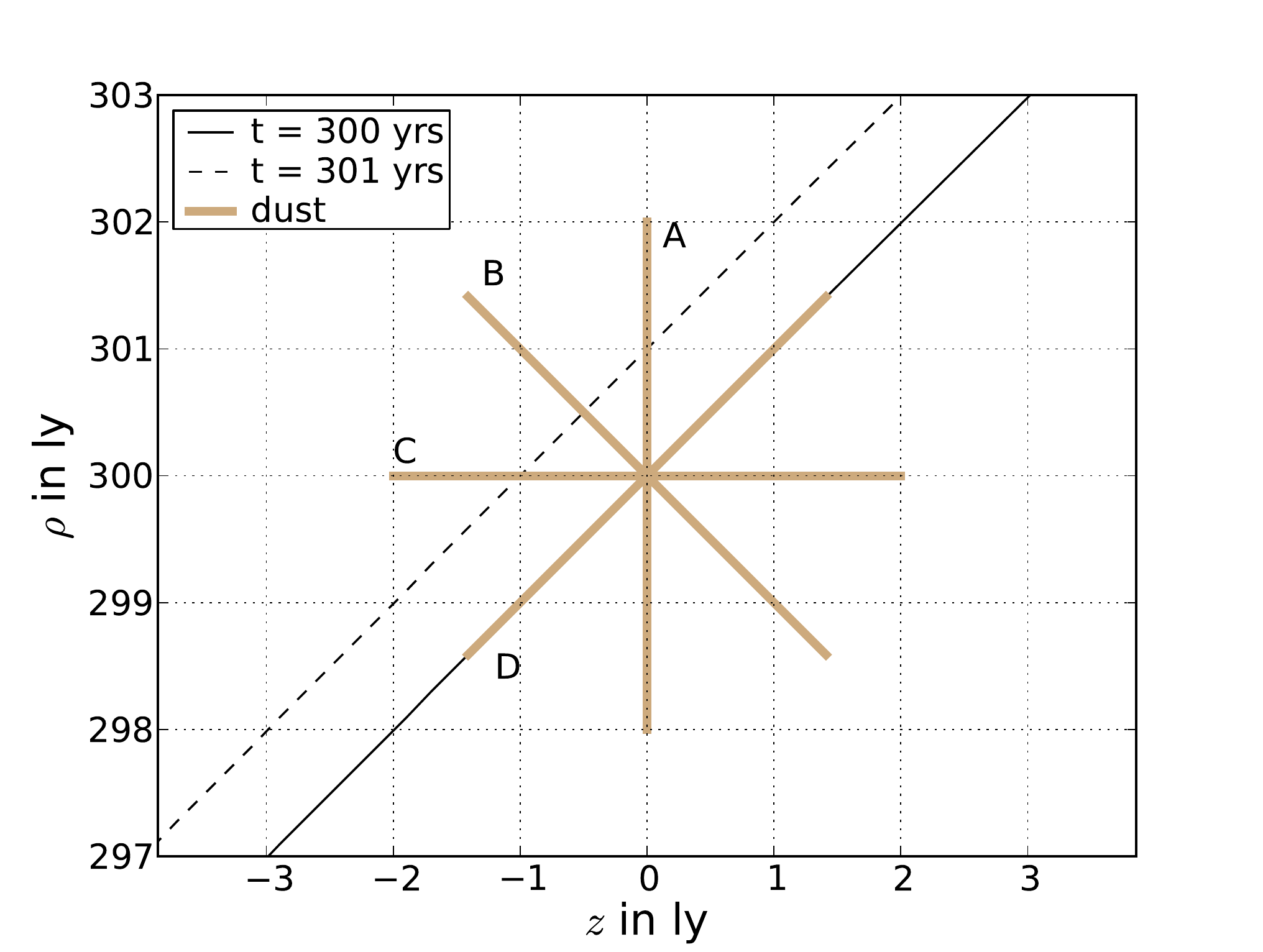}%
\caption[]{ The ellipses of equal arrival time defined
by Equation~\ref{eq:le} are shown as the solid and dashed black line
for $t=300$ and $t=301$~ly, respectively. The dust sheets $A$, $B$, $C$,
and $D$ have inclinations of $\alpha=0$, 45, 90, and 135 degrees,
respectively. LEs scattering off dust sheets $A$, $B$, $C$,
and $D$ have apparent motions of 1.0$c$, 0.5$c$, 0.0$c$, and $\infty$,
respectively. Here, $c$ is the speed of light.
\label{fig:apparentmotion}}
\end{figure}

Figure~\ref{fig:apparentmotion} illustrates how different dust sheet
inclinations produce different apparent LE motions. We define the
inclination angle, $\alpha$, as the angle of the dust sheet with
respect to the $\rho$ axis, where positive angles go from the positive
$\rho$ axis towards the negative $z$ axis.  The ellipses of equal
arrival time defined by Equation~\ref{eq:le} are shown as the solid
and dashed black line for $t=300$ and $t=301$~ly, respectively. The dust
sheets $A$, $B$, $C$, and $D$ have inclinations of $\alpha=0$, 45, 90,
and 135 degrees, respectively, crossing the point
$(z,\rho)=(0,300)$~ly.  LEs scattering off dust sheets $A$, $B$, $C$,
and $D$ have apparent motions of 1.0$c$, 0.5$c$, 0.0$c$, and $\infty$,
respectively. Here, $c$ is the speed of light. One can see that for
dust inclinations in Figure~\ref{fig:apparentmotion} where
$-45\leq\alpha<0$, the apparent motion of the LE is superluminal.
Theoretically, for each apparent motion there exists one unique dust
filament inclination in a 180 degree range that produces exactly that
apparent motion.
For a LE at a given position, there exists degeneracy between the time
$t$ since explosion, the dust inclination, and the apparent motion. In
order to determine one of these parameters, the other two parameters
must be known. However, the inclination of the dust filament is seldom
known. A solution is to marginalize over the dust inclination, showing
that the apparent motion associated with a dust filament with an
inclination perpendicular to the LE ellipsoid is a very good
approximation of the expectation value (see Rest et~al., in prep., for
a detailed derivation). Note that this marginalization assumes that
the inclination of the dust is random, which might not be true due to
detection biases. As an example, a common misconception is that for
$z=0.0$, i.e. the dust is in the plane of the sky, the apparent motion
of LEs is $c$, implicitly assuming that the dust inclination is
$\alpha=0.0\arcdeg$. However, the true expectation value is $v=0.5c$
associated with $\alpha=45\arcdeg$. As we will see below, the
inclination is an important physical property of the scattering dust
that not only affects the apparent motion of the LE, but the observed
flux profile and spectrum of the LE as well.

\subsection{Geometric Distance}
\label{sec:distance}

Under suitable circumstances, LEs provide a unique opportunity
to determine the geometric distance to its source object. For a given
light echo at an angular distance to the source event with known event
time, there is still a degeneracy between distance $D$ (source to
observer) and the distance $z$ (scattering dust to the source event
along the line of sight). This degeneracy gets broken if $z$ can be
estimated with some additional information or assumptions.

One of the easiest assumptions to make is $z=0$, or that the
apparent motion of the light echo is the speed of light.  However, if
these assumptions are not based on any additional evidence, the
systematics can be very large and consequently can lead to incorrect
distances. Examples for this are discussed for V838 Mon
(Section~\ref{sec:V838mon}).

If the geometrical shape of the scattering dust can be
constrained, the degeneracy can also be broken. This has been done in
the case of SN~1987A where \cite{Panagia91} determines the distance to
the LMC to a few percent by assuming that its circumstellar ring is
circular (Note, however, that they use gaseous emission lines, and not
scattered light echoes, which light travel delays follow similar
equations).  \cite{Gould94,Gould98} show that the distance changes
only on the percent level if the ring is moderately elliptical.

Another complication is added if the source of the light
echoes has recurrent maxima. In that case it is difficult to determine
which light echo corresponds to which maximum. If the wrong association
is done, catastrophic errors can be the result, and therefore extra
care needs to be applied when determining these associations. We
discuss two cases in Section~\ref{sec:CrA} and \ref{sec:RSpup}.

One of the most accurate and robust ways of determining the
distance with light echoes is using the polarization of the scattered
light.  The scattering polarizes the light, where the
polarization of this scattered light will vary strongly with the
scattering angle between the source and the observer \citep{Draine03}.
For normal scattering cross sections, the angle of maximum
polarization, $\psi_{pol}$, is close to $90\arcdeg$ and depends only
slightly on dust properties like type and grain size distribution.  If
the time since the source event, $t$, is known and
$\psi_{pol}=90\arcdeg$, the distance to the source event can be
calculated with $ct=D \tan(\gamma) $, where $\gamma$ is the angular
separation of the maximum polarization to the source event.  In the
event that $\psi_{pol}$ is not exactly $90\arcdeg$, but is known, the
method can still be used and the corrected distance is
\begin{eqnarray}
D' = D \times \frac{\cos(\psi_{pol})+1}{\sin(\psi_{pol})}.
\label{eq:distance}
\end{eqnarray}
We refer the reader to \citet{Sparks94,Sparks96} who first
investigated how this method can be used to determine distances.

Since the maximum polarization angle $\psi_{pol}$ depends on the type
of scattering dust (usually unknown), systematic uncertainties on the
order of 5\%-10\% are introduced into these distance measurements
\cite[see Fig. 5 of][]{Draine03}. For example, at a wavelength of
$6165 $~\AA, $\psi_{pol}$ can range from $~90\arcdeg$ to $~97\arcdeg$
for SMC bar dust and Galactic dust, respectively.  Therefore by
determining $\psi_{pol}$ at different wavelengths we can verify if the
polarization is consistent with the standard Galactic $R_V=3.1$ dust
model.  If the dust is indeed consistent with $R_V=3.1$, then the
theoretical prediction for $\psi_{pol}$ is well constrained and the
systematics will be lowered to a few percent.

The most important ingredient for this method is to have LEs
with scattering angles spread around $90 \arcdeg$, which corresponds
to having LEs with distances spread around $z=0$.  
The best example where this method has been applied is V838
Mon, as discussed in Section~\ref{sec:V838mon}, which has an extensive
light echo system spanning a wide range in $z$.  However, not all
objects with LEs have such extensive light echo systems, but Tycho
\citep{Rest08b}, Cas~A \citep{Rest08b}, and most recently
$\eta$~Carinae \citep{Rest12_etac} might have a suitable number of LEs
that cover the required range of angular separations from the source
event for this method to work.

\subsection{Light Echo Profile \label{sec:leprofile}}

The flux profile of a LE is the slice through the LE along the
$\rho$ axis pointing toward the source event. This profile is actually
the lightcurve of the source event stretched or compressed depending
on the inclination, $\alpha$, of the scattering dust filament, and
convolved with the effects of the dust width, $\sigma_d$, and the
seeing \citep{Rest11_leprofile}.

\begin{figure*}
  \includegraphics[width=460pt]{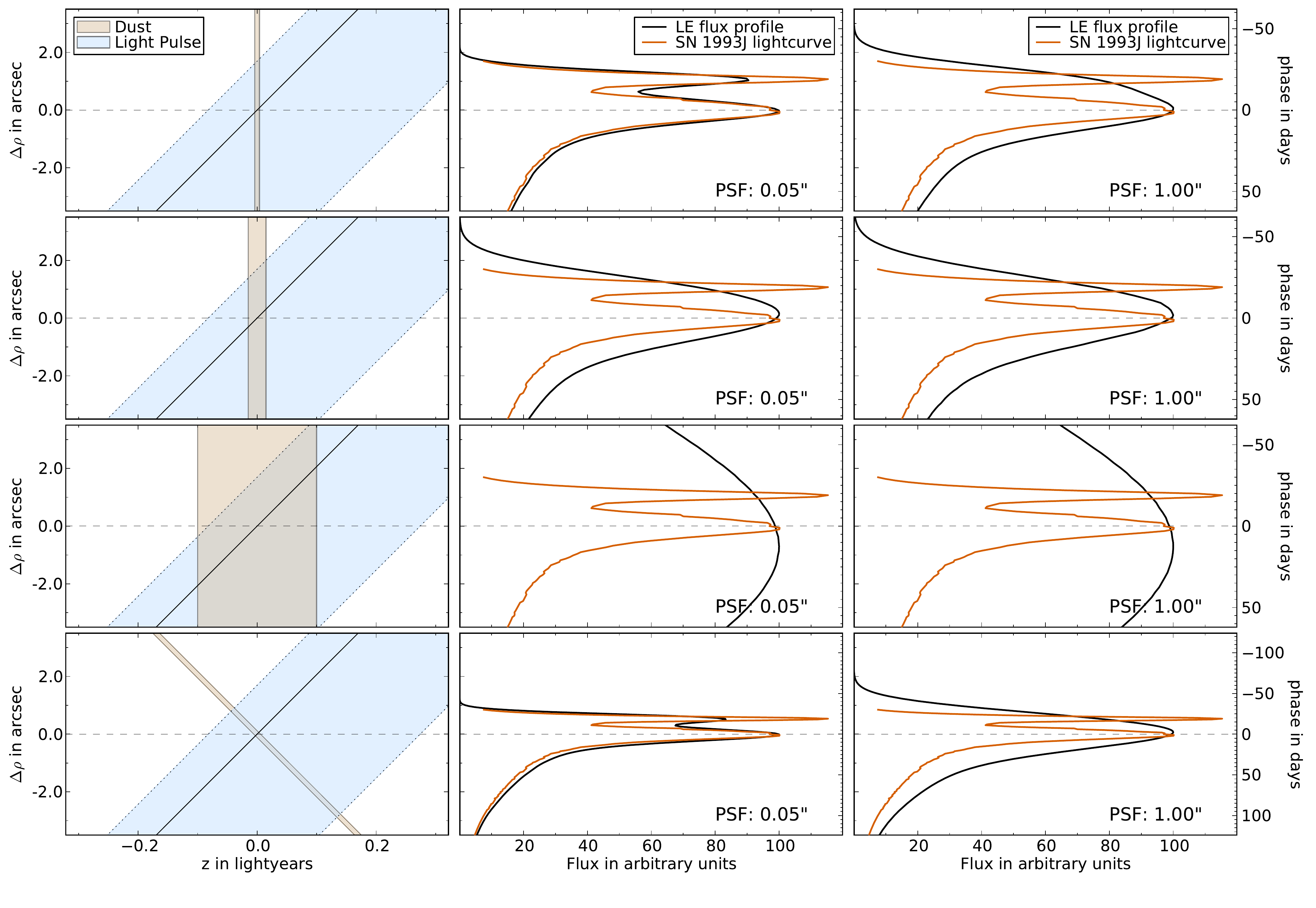}%
   \caption[]{
The blue-shaded area indicates a 130 day long event similar to
SN~1993J with a fast rise, a peak after 30 days, and a slow
decline. The scattering dust filament (brown shaded area) is located
at $(z,\rho) = (0,300)$~ly with dust widths $\sigma_d$ of 0.008~ly,
0.03~ly, 0.2~ly, and 0.008~ly, from top to bottom, respectively. All dust
filaments have an inclination of $\alpha=0\arcdeg$, with the exception of
the bottom panel which has an inclination of $\alpha=45\arcdeg$.  The
corresponding LE flux profiles are shown in the middle and right
panels for PSF sizes of 0.05\arcsec~and 1.0\arcsec, respectively. For
an infinitely thin dust sheet and an infinitely small PSF, the flux
profile is just the projected light curve, as indicated by the red
line. The phase of this projected light curve is shown on the right
y-axis.
\label{fig:widthseeingincl}}
\end{figure*}

As in \citet{Rest11_leprofile}, we illustrate this stretching and
compressing effect using a realistic LE scenario similar to that
observed in the Cas A echo system. We consider an echo originating
from an event in the Galaxy 300 years ago at a distance of 10000
light-years. We assume the outburst event was similar to the observed
outburst of SN~1993J. The double-peaked nature of SN~1993J's
lightcurve makes for a particularly good example for illustrating the
observational effects of the scattering dust and the PSF size. The
left panels in Figure~\ref{fig:widthseeingincl} show scattering dust
filaments (brown-shaded area) of different widths and
inclinations. The blue-shaded area indicates a 130 day long event
similar to SN~1993J with a fast rise, a peak after 30 days, and a
slow decline. Note that the beginning of the outburst is observed at
the largest distance away from the source on the sky, not the closest.
The scattering dust filaments are located at $(z,\rho) = (0,300)$~ly
with inclinations of $\alpha = (0,0,0,45)$\arcdeg~and widths of
$\sigma_d = (0.008,0.03,0.2,0.008)$~ly from top to bottom,
respectively. We project the coordinate $\rho$ onto the sky and show
on the y-axis the angular difference in $\rho$ to the peak in arcsec.

The middle and right panels of Figure~\ref{fig:widthseeingincl} show
the LE flux profiles (black lines) for the different dust
configurations and PSF sizes. For an infinitely thin dust sheet and an
infinitely small PSF, the flux profile is just the projected light
curve, as indicated by the red line. For very thin
dust filaments ($\sigma_d=0.008$~ly) with $\alpha=0$ and {\it
HST}-like PSF size, the observed LE profile still shows the signature
of the double peak in the lightcurve (upper middle panel). For
ground-based seeing conditions (upper right panel), however, the two
peaks are smeared out and indistinguishable. Similarly, if the dust
width increases, more and more of the original lightcurve shape is
lost ($\sigma_d=0.03$~ly, middle panel in the 2nd row), until not much
of the original shape is left ($\sigma_d=0.2$~ly, middle panel in the
3rd row). Note that the effects of dust width and PSF size are somewhat
degenerate.

The effect of the dust filament inclination is illustrated in the
bottom panel, where the LE flux profile is squeezed by a factor of two
compared to the profile shown in the top row due to the inclination of
the dust filament of $\alpha=45\arcdeg$. Fortunately, the dust
filament inclination is straightforward to determine from the apparent
motion if the date of and distance to the source event is known
\citep{Rest11_leprofile}.  A more detailed discussion of the interplay
between dust width, dust inclination, and seeing can be found in
\citet{Rest11_leprofile}. Such accurate modeling of the dust is
essential when looking for small differences in spectral features,
e.g., when comparing spectra of LEs from the same object as done in
\citet{Rest11_casaspec}.

\subsection{Light Echo Spectroscopy\label{sec:lespec}}

While imaging LEs can reveal the nature of the material around an
outburst event, and possibly the distance to the event (as in
Section~\ref{sec:distance}), it is difficult to further our
understanding of the outburst itself through imaging. Spectroscopy of
LEs, however, allows the outburst to be studied in detail after it has
already faded.  The first time a spectrum of a LE was taken was in
1902 of Nova Persei in a 35~hour effort that can only be called
heroic. It confirmed that the nebulous moving features seen around
Nova Persei were indeed its echoes.  \citet{Gouiffes88} and
\citet{Suntzeff88} obtained spectra of the inner and outer ring of
SN~1987A's light echoes, and compared to averaged spectra from SN
1987A.  They found the LE spectra were most similar to those of the SN
near maximum light. Serendipitously, \cite{Schmidt94} found that 750
days after maximum, the spectrum of SN~1991T showed its light echo
spectrum reflected from foreground dust superimposed on the late-time
nebular spectrum.  Since 2008 with the work of \citet{Rest08a},
studies using LE spectroscopy have shown it to be a powerful tool to
spectroscopically type ancient and historic SNe.

Initial studies assumed that an observed LE spectrum represented a
lightcurve-weighted integration of the SN spectra at individual
epochs \citep[e.g.,][]{Rest08a,Krause08a}. In most cases, this is
adequate for an approximate spectral identification and classification
of the source event. However, \cite{Rest11_leprofile} showed that an
observed LE spectrum represents an integration weighted with an
\emph{effective} lightcurve. The effective lightcurve can be
constructed by applying a window function to the originally observed
lightcurve. Modeling both astrophysical (dust inclination, scattering,
and reddening) and observational (seeing and slit width) effects for a
given LE are required to determine the correct window function and
therefore correctly interpret the observed (integrated) LE spectrum.
This is illustrated in Figure~\ref{fig:widthseeingincl}: If a slit is
placed on the LE profile, in the cases where the width of the dust
filament is not very large, the window function is narrow and
so the slit probes only a limited part of the source event epochs
(1st and second row of panels from the top).  Only for very thick dust
filaments (3rd row of panels from the top) is the window
function sufficiently wide such that the spectrum probes all or
nearly all epochs of the source event.

An important part of determining the correct window function is
determining if late-time spectral features are included in the observed
LE spectrum. Naively, one might expect these late-time features are
not significant since the surface brightness of the source event already
significantly declined. However, at late times when
the spectrum is nebular, the observed flux is typically concentrated
into a few strong lines. We demonstrate the importance of this
late-time contribution using LE spectroscopy from SN~1987A
\citep{Rest11_leprofile}. The SN~1987A system is the ideal testbed for
LE spectroscopy, where the high-flux LEs can be paired with the
complete spectral and photometric history of the SN as it was
originally observed.

\begin{figure*}
   \includegraphics[width=460pt]{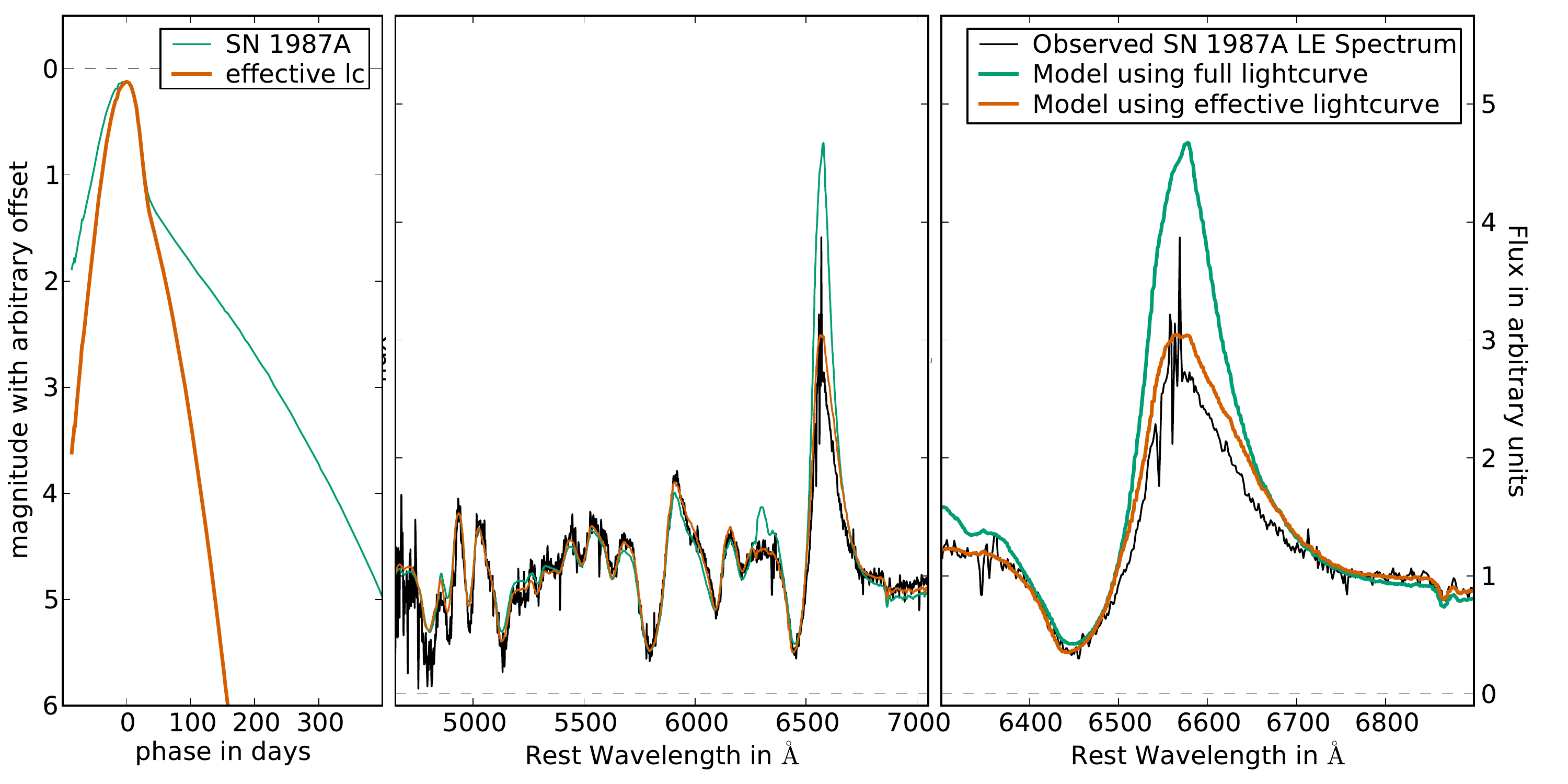} 
   \caption[]{
Analysis for SN~1987A LE spectrum taken using the Gemini
Multi-Object Spectrograph on Gemini-South. Left: Original lightcurve of
SN~1987A compared with the effective lightcurve associated with the observed
LE. Middle: Observed LE spectrum of SN~1987A (black line) compared with the
modeled spectra created by integrating the original outburst spectra of
SN~1987A with the two lightcurves shown in the left panel. Right: Closeup of
H$\alpha$ line, showing that only the modeled \emph{effective}
lightcurve can correctly account for the relative strength of the
H$\alpha$ line in the observed spectrum.
\label{fig:87a}}
\end{figure*}

Figure~\ref{fig:87a} shows a modified and updated version of
Figure~18~and~19 in \cite{Rest11_leprofile}, which compares a SN~1987A
LE spectrum to its model.  The left panel shows the lightcurve of
SN~1987A \citep{Hamuy88,Suntzeff88b, Hamuy90} and the effective
lightcurve resulting from multiplying the lightcurve with the
appropriate window function.  The window function is determined from
the best-fitting modeling of the LE profile, taking dust width and
inclination, seeing and slit width into account. The necessity of this
modeling is clear when comparing model spectra (colored lines) to the
observed LE spectrum (black line) in the middle and right panels of
Figure~\ref{fig:87a}. The model spectra shown here are integrations of
the originally observed spectra of SN~1987A \citep{Menzies87,
Catchpole87,Catchpole88, Whitelock88,
Catchpole89,Whitelock89,CTIO_87aspec1,CTIO_87aspec2} weighted with the
effective (orange line) and full (cyan line) lightcurves shown in the
left panel.  Here, modeling the correct window function limits the
contribution of late-time spectra into the final integrated model
spectrum. In the case of SN~1987A, the late-time spectra show strong
nebular emission flux in H$\alpha$. The H$\alpha$ line is shown in the
right panel of Figure~\ref{fig:87a}, where it is fit well with the
effective lightcurve model.  Incorrectly modeling the observed LE
spectrum using the full lightcurve of SN~1987A results in an emission
component to the H$\alpha$ line that is far stronger than actually
observed in the LE spectrum. It is clear from Figure~\ref{fig:87a}
that detailed consideration of the scattering dust and observational
effects are critical in correctly interpreting LE spectra.
Examples of LE spectroscopy are
discussed in Section~\ref{sec:etac},~\ref{sec:lmc},~\ref{sec:casa},~and~\ref{sec:tycho}.

\subsection{3D Spectroscopy\label{sec:3Dspec}}

\begin{figure*}
   \includegraphics[width=460pt]{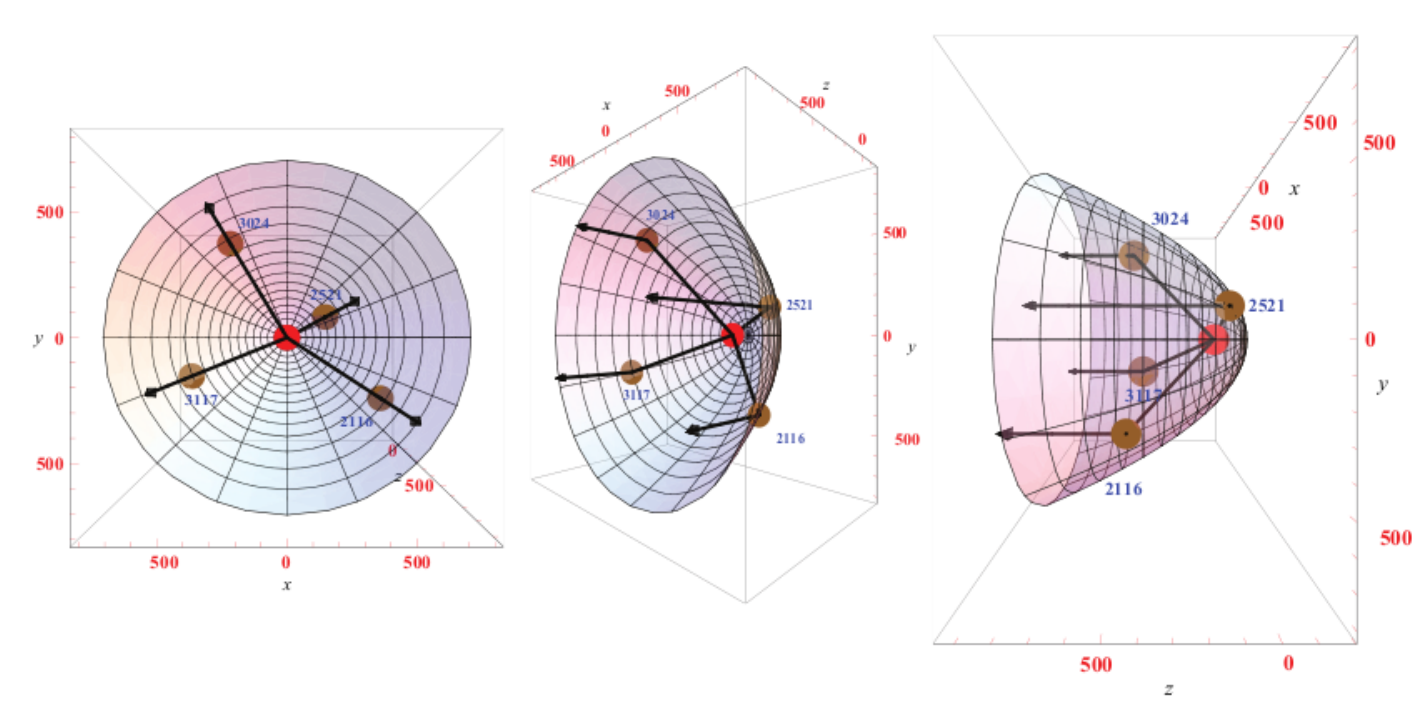} 
   \caption[]{3D illustration of Cas~A LEs. North is toward
the positive-$y$ axis (up), east is toward the negative-$x$ axis
(left), and the positive-$z$ axis points toward the observer with the
origin at the Cas~A SNR (red filled circle). The four brown circles
indicate the scattering dust of LEs discovered by
\cite{Rest08b}. The black lines show the path of the light scattering
from the LE-producing dust concentrations.
\label{fig:CasA_3Dspecview}}
\end{figure*}

LEs offer the exciting opportunity to obtain spectra of the original
source event from {\it different} lines of sight (LoSs).
Fig.~\ref{fig:CasA_3Dspecview} illustrates how LEs from Cas~A's SN,
which are scattered by different dust structures, probe the SN from
different directions.  Observing these arcs is equivalent to observing
different hemispheres of the photosphere, providing direct
observations of potential asymmetries in the source event.
\citet{Smith03_eta} first applied this technique, observing the
$\eta$~Carinae central star from different directions using spectra of
the reflection nebula.

The signatures of asymmetries in spectra of the same events are much
more subtle than when doing spectral classification by comparing
different events. Therefore it is essential that both astrophysical
(dust inclination, scattering, and reddening) and observational
(seeing and slit width) effects are taken into account accurately, as
described in the previous Section~\ref{sec:lespec} and in more detail
in \citet{Rest11_leprofile}. An example of 3D-spectroscopy is
discussed in Section~\ref{sec:casa}.

\subsection{Spectroscopic Time Series\label{sec:spectimeseries}}

One as-yet unrealized opportunity provided by SN LEs is the ability to
obtain a spectral time series of the source event of LEs using very
thin scattering dust filaments, as suggested by
\cite{Rest11_leprofile}. As discussed in Section~\ref{sec:leprofile},
the LE profile is simply the spatially projected light curve of the
source event, convolved with the finite thickness of the scattering
dust filament and the finite size of the PSF.  The spatial extent of
the projected lightcurve is determined by the inclination of the dust
filament, in combination with the duration of the event. For short
events like SNe, and typical dust configurations in our Galaxy, the
spatial extent is on the order of arcseconds (see
Figure~\ref{fig:widthseeingincl}). This projected lightcurve is then
smeared due to the thickness of the dust filament, in combination with
the PSF size. Whether it is possible to temporally resolve the SN
spectra depends on the combination of the above parameters

Figure~\ref{fig:widthseeingincl} illustrates this: if the dust
filament is very thin, for example as observed for Cas~A LEs
\citep{Rest11_leprofile}, the resulting window function can have a
width of $\approx$10 days. In the example of SN~1993J, a SN~IIb
similar to Cas~A, it would allow us to observe the spectrum of the
shock breakout with an instrument with HST-like PSF-size (top and
bottom middle panel). An enormous advantage of using LEs for observing
the shock break-out is that there is no urgency to obtain
observations.  Every pixel is a data point, and pixels before the edge
of the LE are equivalent to closely spaced non-detections in a SN
search. With this method, getting a spectrum at early epochs is as
easy (or difficult) as getting a spectrum at late epochs, which is in
stark contrast to SN searches for which it is difficult to get spectra
at early epochs due to the lag time between when the image is taken,
the SN is discovered, and follow-up is triggered.

Spectra at early epochs of a SN or other transients are especially
valuable scientifically, since they contain many signatures of the
original explosion and also the progenitor. For example, with the
shock break-out, the duration, luminosity, and SED depends on a
handful of parameters such as the presence of a stellar wind and the
ejecta mass \citep{Matzner99}, but is most dependent on the progenitor
radius \citep{Calzavara04}.  The peak brightness of the subsequent
fading is dependent on the ejecta mass, kinetic energy, and progenitor
radius, while the timescale for the fading is proportional to both the
radius of the photosphere and its temperature \citep{Waxman07}.  The
ability to temporally resolve the SN spectra allows one to follow the
evolution of the explosion, e.g., the evolution of the velocity
gradients of spectral features and the abundances of elements in the
ejecta \citep{Rest11_leprofile}.

A particularly promising case for this technique is if the duration of
the source event is very long, since the spatial extent of the
projected lightcurve is proportional to its duration. For example,
since the Great Eruption of $\eta$~Carinae spanned more than a decade,
it is now possible to obtain LE spectra of it from different epochs
\citep{Rest12_etac}, which are only marginally affected by dust width
and PSF-size of the observations.

\section{Light Echo Case Studies}

In recent years, new LEs have been discovered for a variety
of events, and new techniques have been used to utilize them.  In this
section we discuss some of the most prominent LE systems and
note the achievements, difficulties, and shortcomings of their
analysis (see also Table~\ref{table:overview}).

\subsection{S~CrA \& R~CrA\label{sec:CrA}}

In a recent paper, \citet{Ortiz10} presented the results of their
imaging of LEs around two young stellar objects, the T Tauri
star S~CrA, and the Herbig Ae/Be star R~CrA. These classes of variable
stars are known for irregular (non-repeating) variability. The authors
draw attention to the discovery of the reflection nebulosity and its
variation by \citet{Hubble22} and \citet{Lightfoot89} although the
early work did not interpret their findings as LEs.

The observational material available to \citet{Ortiz10} for its
analysis of S~CrA consisted of I-band photometry from the All Sky
Automated Survey \citep{Pojmanski97} spanning approximately 150 days
and nine images spanning 93 days. They adopted a model of a
spherically-symmetric shell of dust centered on the star and then
associated the four brightness peaks from the ASAS time-series with
the observed LE features from their imaging. Based on their
assumptions they arrived at a distance to S~CrA of $128 \pm 16$~pc, a
distance between the star and dust shell of $10^4$~AU. \citet{Ortiz10}
speculated that the dust shell could be similar in nature to the Oort
cloud surrounding our Sun.

The major uncertainty in distance estimates using LEs is the
scattering geometry which is unknown without polarization measurements.
Distances derived with polarimetry are also subject to systematic
errors as we shall describe in Section~\ref{sec:RSpup} and
\ref{sec:V838mon}, even when the dust distribution is described correctly. The
analysis of S~CrA LEs has several
additional flaws. First, given four brightness peaks in four months
and nine images spanning three months, it is very likely that there is
at least one repeating LE - that is, a LE appearing
twice caused by two different peaks scattering off the same dust
filament.  Second, the brightness differences between the peaks do not
seem to be large enough to be able to explain this.  Third, there is a
large analysis degeneracy between the distance to the star, the times
since peak brightness(es), and the distance from the dust to the star. 
This degeneracy has not yet been fully explored.  Finally, the time series
itself is noisy and the identification of brightness peaks is not
unambiguous. While it may be possible to determine a useful geometric
distance to S~CrA using LE methods and polarimetry, we
conclude that the existing distance estimate is flawed.

\subsection{RS Pup\label{sec:RSpup}}
 
RS Puppis is one of the longest-period and consequently most luminous 
classical Cepheids known in our Galaxy. As such it is one of the Galactic 
counterparts of the bright Cepheids used to determine distances to 
the most distant galaxies amenable to distance estimation with this
standard candle.  Since Cepheids are believed to be a high-reliability 
distance indicator, determining an accurate, independent ``geometric''
distance to RS Puppis, which at a distance of approximately 2.0~kpc
cannot be determined by trigonometric parallax with current 
instrumentation, would be highly desirable.

The reflection nebula of RS Puppis was discovered by \citet{Westerlund61}, 
who remarked on the possibility of LEs due to the variable nature 
of the star. \citet{Havlen72} used a series of photographic images
to demonstrate that features within the nebula show light variations at 
the Cepheid's pulsation period and argued that these ``echoes'' could be 
used to derive a geometric distance, provided that the phase angle of
the scattered light was known. More recently, \citet{Kervella08} 
obtained an observing season's worth of high-angular resolution CCD images
and used phase lag measurements of the RS Puppis LEs to estimate 
a geometric distance of $1992 \pm 28$~pc, which if correct,
would be by far the most accurate distance to a Cepheid. It would also be a
distance unaffected by uncertainty in reddening.

However, \citet{Bond09} showed that the analysis by \citet{Kervella08}
has serious flaws. They pointed out that the implicit assumption by
\citet{Kervella08} that the scattering dust lies, on average, in the
plane of the sky is not valid since the scattering efficiency
strongly favors forward scattering and thus significantly biases the
results. In addition, \citet{Bond09} showed that the \citet{Kervella08}
results have low statistical significance and argued that with
polarization measurements, as demonstrated with V838 Mon by
\cite{Sparks08}, the true geometric distance might yet be 
determined with high precision using this technique.

\subsection{V838 Mon\label{sec:V838mon}}

Some of the most spectacular LEs \citep{Henden02,Bond03} appeared when
V838 Monocerotis went through an explosion in 2002 \citep{Brown02}, an
outburst so energetic that it became one of the brightest stars in the
Local Group for a few weeks in 2002 at $M_V=-10$, gaining 9 magnitudes
in the {\it V} band from its typical quiescent brightness. It ejected
so much debris that the material is still not optically thin and is
now thought of as a member of a new class of variables, the so-called
intermediate-luminosity red transients.

The echoes around this star were subsequently used to determine its
distance. The first attempts using LE expansion measurements greatly
underestimated the distance to V838 Mon to be $<1$~kpc
\citep{Munari02a,Kimeswenger02}, because they were based on the
assumption that the LE apparent motion is at the speed of light.  More
realistic estimates were first done using polarization measurements
from HST imaging, setting a lower limit of $\le 6$~kpc
\citep{Bond03}. A detailed construction of the 3D map of the
scattering dust by \cite{Tylenda04} provided a similar limit of $\le
5$~kpc, and also indicated that the scattering dust is of interstellar
and not circumstellar nature. Assuming that the dust is in a thin
sheet, \cite{Crause05} finds a distance of $8.9 \pm 1.6$~kpc.  The
most accurate distance determination to date is provided by
\citet{Sparks08}, whose detailed analysis of the LE polarization and
expansion speed yield a distance of $6.1 \pm 0.6$~kpc, in very good
agreement with other methods \citep[e.g.,][]{Asfar06}.

The LEs also show a remarkable feature caused by a double helix in the
scattering dust filaments, which point almost radially towards V838
Mon \citep{Carlqvist05}. A possible mechanism for the formation of
such a complex structure is the twisting of outflowing material by a
magnetic field, as suggested by \citet{Carlqvist05}.

V838 Mon also shows IR echoes \citep{Banerjee06}, which are spatially
coincident with the scattered LEs. The large mass of the dust
($>0.2$~$\Msun$) makes it unlikely that the dust is circumstellar and
from a previous mass-loss episode, supporting the hypothesis that the
dust causing both the scattered and IR echoes is interstellar
\citep{Banerjee06}.

\subsection{SN~1987A\label{sec:87A}}

Shortly after SN~1987A exploded, spectacular LEs scattered off
interstellar dust at distances of $\sim$100 and $\sim$400~pc
\citep{Crotts88,Suntzeff88,Gouiffes88,Couch90}. Later, LEs scattered
off circumstellar dust were also discovered
\citep{Crotts89,Bond89}. These LEs have revealed valuable information
about the circumstellar environment around SN~1987A
\citep{Sugerman05a,Sugerman05b}, the interstellar medium (ISM) of the
LMC \citep{Xu94,Xu95,Xu99}, and about the asymmetry of SN~1987A itself
(Sinnott~et~al., in prep).

\subsubsection{SN~1987A light echoes scattered off circumstellar dust}

\citet{Crotts91} uses the light echoes within 10\arcsec~of SN~1987A to
determine the density and 3D structure of the scattering dust,
constraining the mass loss of SN~1987A's progenitor. They found that
the scattering dust is consistent with being from mass lost at a
constant rate from a red supergiant atmosphere. Later observations
revealed a double-lobed nebula, with a waist that is nearly coincident
with the elliptical circumstellar ring seen in atomic recombination
lines \citep{Crotts95}.

\citet{Sugerman05a} uses LEs detected within 30" of SN~1987A to
construct the most detailed 3D-model of the circumstellar environment.
They found that a richly structured bipolar nebula surrounds SN~1987A,
with an outer, double-lobed ``peanut'' extending 28 lt-yr along the
poles, which is believed to be the contact discontinuity between the
red supergiant and main-sequence winds. The waist of this peanut is
the Napoleon's Hat, which was previously thought to be an independent
structure. The innermost circumstellar material is in the form of a
cylindrical hourglass, 1 lt-yr in radius and 4 lt-yr long, connected
to the peanut by a thick equatorial disk \citep{Sugerman05a}.  With
this 3D-model of the scattering dust the progenitor's mass-loss
history can be reconstructed \citep{Sugerman05b}.  They found from the
echo fluxes that from the interior hourglass to the bipolar lobes, the
gas density drops from 1-3 to 0.03~cm$^{-3}$, while the maximum
dust-grain size increases from 0.2 to 2~\micron, and the
silicate-to-carbonaceous dust ratio decreases, resulting in a total
mass of \about1.7~$\Msun$.  The studies of
\citet{Sugerman05a,Sugerman05b} are the most detailed
three-dimensional circumstellar dust reconstruction of any stellar
object to date.

\subsubsection{SN~1987A light echoes scattered off interstellar dust \label{87aIMSLEs}}

It is of special interest to understand the structure of the ISM near
SN~1987A in the LMC. This is a highly active starburst region with
massive stars emitting strong stellar winds. The interaction of these
high-energy outflows with the filamentary and spherical structures of
the ISM produce bubbles and super-bubbles in the region which trigger
additional star formation \citep[e.g.,][]{Walborn99}. LEs from
SN~1987A allow the structure of these ISM supper-bubbles to be traced
out in three dimensions.

There are two nearly complete LE rings observed around SN~1987A,
caused by two dust sheets roughly perpendicular to the line-of-sight
$\sim$100 and $\sim$400~pc in front of the SN
\citep{Crotts88,Suntzeff88,Gouiffes88,Couch90}.  \citet{Xu94} found
two additional LEs at a much larger distance of $\sim$1000~pc in front
of SN~1987A. All of the above LEs were used by \citet{Xu95} to
construct a detailed 3D-map of the ISM scattering dust.  Comparing
with H$\alpha$ gas maps, \citet{Xu95} associate the ring at
$\sim$400~pc with reflection from the boundary of the super-bubble
N157C. They found the curvature of this dust complex to coincide with
LH~90, the source of the super-bubble. Additionally, the southeastern
arc at a distance of \about1000~pc was found to align
with a massive H$\alpha$ filament discovered by
\citet{Meaburn84}. They speculate that these two structures are the
near and far sides of a giant super-bubble with a diameter of
$\sim$600~pc, which itself may have resulted from the merger of
several smaller super-bubbles.

\subsubsection{Asymmetry of SN~1987A}

The two main SN~1987A LE rings scattered off interstellar dust (see
Section~\ref{87aIMSLEs}) provide an excellent opportunity to search
for spectroscopic asymmetries in SN~1987A using the 3D spectroscopy
technique detailed in Section~\ref{sec:3Dspec}.  These LEs are very
bright, allowing the observation of high signal-to-noise ratio
spectra. A wealth of previous imaging of the LEs also exists, making
the observed apparent motion and hence dust inclination well
known. Most importantly, however, the original spectral and
photometric history of SN~1987A are both well known, resulting in no
ambiguity when modeling the LE spectra (Section~\ref{sec:lespec}).

Sinnott~et~al. (in prep.) have obtained spectra of 14 LEs of the inner
echo ring of SN~1987A.  In order to reduce the effects of nebular and
stellar contamination, they have also obtained follow-up sky-only
observations at the same position after the LE moved away using the
same instrument configuration for the purpose of difference
spectroscopy. In addition to providing multiple viewing angles onto
SN~1987A, the LEs which scatter off slightly different dust sheets
while probing essentially the same viewing angle allows for a direct
observational test of the LE spectra modeling techniques described in
Section~\ref{sec:lespec}.  After properly taking the dust filament
extent and inclination into account (see Figure~\ref{fig:87a}), they
find evidence for asymmetry along a north-east to south-west axis,
with little signs of asymmetry seen in the equatorial east-west
directions. The details of this asymmetry will appear in
Sinnott~et~al. (in prep.).

\subsection{The 19th-Century Great Eruption of
$\eta$~Carinae \label{sec:etac}}

$\eta$~Carinae ($\eta$~Car) is the most massive and most studied star
in our Galaxy. It became the second-brightest star in the sky during
its mid-19th century ``Great Eruption,'' in which it lost more than
10~$M_{\odot}$ \citep{Smith03}.  \citet{Rest12_etac} discovered LEs of
the Great Eruption, and subsequent spectroscopic follow-up revealed
that its spectral type is most similar to those of G-type supergiants,
rather than to F-type or earlier like typical luminous blue variable
(LBV) outbursts. This raises doubts that traditional models involving
opaque winds explaining LBV outbursts can fully explain the Great
Eruption. The absorption lines of the LE spectra indicate the emitting
photosphere has an ejection velocity of $\sim -200$~km~s$^{-1}$ from a
viewing angle perpendicular to the principal axis of the Homunculus
Nebula \citep{Rest12_etac}. This is in agreement with velocities
predicted by \citet{Smith06}.  The $\eta$~Car LE spectrum also shows a
strong asymmetry in the Ca~II IR triplet, extending to a velocity of
$-$850~km~s$^{-1}$ \citep{Rest12_etac}.

In recent years, some extragalactic non-SN transients, (dubbed ``SN
impostors''), have been interpreted as analogues of the Great Eruption
of $\eta$~Car \citep[e.g.,][]{Vink09}, even though the Great Eruption
is an extreme case in terms of energy, mass-loss, and duration. It is
nevertheless surprising how different the LE spectra of the Great
Eruption is compared to the SN impostors.  Its spectral type is G2-G6,
significantly later than all other SN impostors at peak, and it does
not show any significant Hydrogen or Ca~II IR triplet emission lines
\citep{Rest12_etac}.

\subsection{Light Echoes of Ancient SNe in the LMC\label{sec:lmc}}

The first LEs of ancient or historic SNe were discovered in the LMC
by \cite{Rest05b}. They found 3 LE complexes associated with LMC SNRs
0509-675, 0519-69.0, and N103B. Using measurements of the LE apparent
motions, they determined the ages of these SNRs to be between
$400-800$~years.

A spectrum of one echo associated with SNR 0509-675 reveals that the
echo light is from the class of over-luminous Type Ia SNe
\citep{Rest08a}, the first time that an ancient SNe got typed based on
its LE spectrum. This result is in excellent agreement with its
classification based on X-ray spectra of the SNR 0509-675
\citep{Hughes95,Badenes08}.

\citet{Hughes95} analyzed X-ray and radio data of the six SNRs in the
LMC which have diameters less than 10~pc and are thus presumably the
youngest in the LMC. They found three SNRs which have element
abundances consistent with nucleosynthesis from Type~II SNe, with the
remaining three consistent with Type~Ia nucleosynthesis.  It should be
noted that the three SNRs classified as Type~Ia are also the SNRs with
known LEs. This classification was confirmed for SNR 0509-675
\citep{Rest08a} and SNRs 0509-675 and 0519-69.0 (Rest~et~al., in
prep.). Only for the by far youngest core-collapse SN, SN~1987A, were
LEs detected. Even though these are low-number statistics, the bias
toward detecting LEs from Type~Ia SNe can be explained by the fact
that core-collapse SNe are typically much fainter than SN~Ia.

\subsection{Cas~A SN\label{sec:casa}}

Cas~A is the youngest (\about 330~years old) core-collapse SNR in our
Galaxy \citep{Stephenson02}. \citet{Krause05} identified features in
{\it Spitzer} Space Telescope IR images that changed with
time. They identified them as IR echoes, which are the result of
dust absorbing the SN light, warming and re-radiating it at longer
wavelengths. However, their main scientific conclusion that the cause
of most if not all of these IR echoes was a series of recent X-ray
outbursts from the compact object in the Cas~A SNR turned out to be
incorrect. They neglected to take into account that the apparent
motion strongly depends on the inclination of the scattering dust
filament \citep{Dwek08} (see also Section~\ref{sec:appmot}). Instead,
the IR echoes are much more likely caused by the intense and short burst
of EUV-UV radiation associated with the shock breakout of the Cas~A
SN itself \citep{Dwek08}. \cite{Kim08} used the IR echoes to reconstruct
the 3D-structure of the absorbing dust filaments.

The first scattered LEs of the Cas~A SN were discovered by
\citet{Rest07, Rest08b}. At the same time, \citet{Krause08a}
spectroscopically followed up one of the IR echoes identified by the
{\it Spitzer} Space Telescope, and found that the Cas~A SN is most
similar to the Type IIb SN~1993J. This implies that the progenitor of
the Cas~A SN was a red super giant that had lost most of its hydrogen
envelope before exploding \citep{Krause08a} .

\citet{Rest11_casaspec} obtained spectra of Cas~A LEs from three
different LEs spatially separated by degrees. This means that each of
the three spectra view the Cas~A SN from different viewpoints (see
Section~\ref{sec:3Dspec}). They accounted for the effects of finite
dust filament extent and inclination, and found that the He~I $\lambda
5876$ and H$\alpha$ features of one LE are blue-shifted by an
additional \about 4000~\kms\ relative to the other two LE spectra and
to the spectra of SN~1993J, indicating that Cas~A was an intrinsically
asymmetric SN. Data of the Cas~A SNR in the X-ray and the optical also
show that there is a dominant Fe-rich outflow in the same direction
\citep{Burrows05, Wheeler08, Delaney10}, in excellent agreement with
the LE data.  This allows for the first time structure in the SNR to
be directly associated with asymmetry observed in the explosion
itself.

\subsection{Tycho's SN\label{sec:tycho}}

Tycho's SN of 1572, one of the last two naked eye SNe in the Galaxy,
was classified as Type Ia based on its observed historical lightcurve
and color evolution \citep{Ruiz04a}. \citet{Badenes06} finds that
modeling the X-ray spectra of the Tycho SNR is most consistent with
Tycho's SN being a normal Type~Ia, the first time that the subtype of
Tycho was determined, even though only indirectly.

Several LE groups of Tycho's SN were discovered by
\citet{Rest07,Rest08b}, which opened the door to spectroscopically
classify this historical SN. \citet{Krause08b} followed up one of
these LEs, and classified it as a normal SN~Ia by comparing it
qualitatively to a small number of SN LE templates constructed from
spectrophotometry of nearby SNe, confirming the classification by
\citet{Badenes06}. It is noteworthy that \citet{Krause08b} did not
quantitatively compare the LE spectrum to a large set of comparison
spectra templates, in particular they only compare the LE spectrum to
a single underluminous SN~Ia. \citet{Rest08a} has shown that even
comparing the high-S/N LE templates to themselves only give conclusive
classifications if a large number of templates are available.

\section{Conclusions}

In the last decade, the number of astrophysical objects with observed
LEs has dramatically increased. Newly-recognized LEs have been found
around extra-galactic SNe as well as LEs from sources in the the Milky
Way and Magellanic Clouds. In particular, LEs from nearby sources have
proven to be fertile grounds for furthering our understanding of
transients and variables. The field has advanced from impressive
geometric reconstructions of circumstellar and interstellar dust
distributions, such as those produced from the study on SN1987A, to
the realization of some of the early promise of improved geometric
distances. The highly-unusual outbursting star V838~Mon had its
distance determined to 10\% based on polarization measurements of the
LEs by determining its geometric distance with an accuracy of better
than 10\% \citep{Sparks08} without any appeal to estimated
luminosities of possible counterparts.

An exciting and promising new phase of SN science was ushered in with
the discovery of whole LE systems around ancient/historic transients
in the LMC and our Galaxy, including Cas~A, Tycho, and $\eta$~Car,
where follow-up spectroscopy has allowed the direct comparison of the
transient with its remnant.  LEs have also allowed now been used, for
the first time, to examine transients from the perspectives of their
reflection nebulae (scattering dust) and have provided observational
spectroscopic constraints on the degree of asymmetry of outbursting
objects.  One as-yet unrealized opportunity provided by LEs is the
ability to obtain spectroscopic time series of events and allow the
reconstruction of the originally-unobserved lightcurve shapes of SNe.

There is a great deal of discovery space left in LE research since
only a small fraction of the sky has been searched for time-variable,
low-surface brightness, non-stellar features. As currently envisaged,
the next generation of wide-field, time-domain surveys like Pan-STARRS
\citep{Kaiser10}, PTF \citep{Rau09}, Skymapper \citep{Keller07}, and
ultimately LSST \citep{Ivezic08} are being designed to study the
celestial sphere based dominantly on the variability of
point-sources. A mode in which whole images of the sky are preserved
with a cadence of weeks or months is likely to reveal an abundance of
additional LE features and the history and perspectives they carry in
their scattered light.

\section*{Acknowledgments} 

We thank M. Bode for bringing the Nova Persei light echo spectrum to
our attention.

\bibliographystyle{fapj}
\bibliography{lerev}

\clearpage
\clearpage
\begin{sidewaystable}[!htbp]
\centering
 \vspace{-3in}
\begin{flushright}
\scriptsize
\tymin=50pt
\begin{tabulary}{\textwidth}{|l|L|L|L|L|L|L|}
\hline
{\bf Associated Source} & {\bf Type} & {\bf Discovery} & {\bf 3D-Structure of
Dust} & {\bf Distances} & {\bf Spectroscopic Classification} & {\bf
3D-Spectroscopy} \\
\hline
RS~Pup & 

Cepheid
& 
 
Reflection nebula discovered by \cite{Westerlund61}. LEs in
reflection nebula discovered by \citet{Havlen72}.
& 

& 
 
Geometric distance was derived without polarization measurements by
\cite{Kervella08}, but \cite{Bond09} shows that analysis is flawed.
& 

& 
 
 \\
\hline
 S~CrA \& R~CrA& 

YSO
& 
 
Discovered by \cite{Ortiz10}.
& 

Scattering dust at $\sim$10000~AU. However, analysis contains flaws and it is
uncertain if this is correct.
& 

Distance of $138\pm16$~pc by \cite{Ortiz10}. However, analysis
contains flaws and it is uncertain if this is correct.
& 

& 
 
\\
\hline
Nova Persei 1901 & 

Nova
& 
 
Discovered by \cite{Ritchey01b,Ritchey01a,Ritchey02}, first
correctly interpreted as such by \cite{Kapteyn02}.
& 

(Some) of the LEs are scattered by dust associated with its planetary nebula
\citep{Seaquist89,Bode04}.
& 
 
& 

35~hour spectrum of nebulosity around Nova Persei shows that it is its
LE, dominated by light from the peak \citep{Perrine03}.
& 
 
 \\
\hline
V838 Mon & 
Red Luminous Transient 
& 
 
Discovered by \cite{Henden02}.
& 

3D maps of the
interstellar dust \citep{Tylenda04}.
& 
 
Distances using angular expansion rates first greatly underestimated
\citep{Munari02a,Kimeswenger02}, improved with better assumptions
\citep{Bond03,Tylenda04,Crause05}, and improved further with
polarization measurements \citep{Sparks08}.
& 

& 
 
 \\
\hline
$\eta$~Car & 

LBV outburst
& 
 
Great Eruption LEs discovered by \cite{Rest12_etac}.
&

&

& 

Great Eruption spectrum is similar to G-type supergiants, and does not
show any evidence of hydrogen or Ca IR triplet emission lines
\citep{Rest12_etac}.
&

\\
\hline
SN~1987A & 

SN II
& 
 
Discovered by  \citep{Crotts88,Suntzeff88}.
& 
3D-maps of circumstellar \citep{Crotts91,Crotts95,Sugerman05a,Sugerman05b} and interstellar dust
\citep{Xu94,Xu95,Xu99}.
& 
 
& 

\citet{Suntzeff88} and \citet{Gouiffes88} obtain the first SN LE spectra of the
outer rings. Show spectra resemble that of maximum light spectrum.
& 
 
Asymmetry detection by Sinnott~et~al., in preparation.
\\
\hline
SNR 0509-675 & 

SN~Ia
& 
 
Discovered by \cite{Rest05b}.
& 

& 
 
& 

Classified as overluminous 91T-like SNe~Ia
with $\dm15<0.9$ \citep{Rest08a}.
&

\\
\hline
SNR 0519-69.0 & 

SN~Ia
& 
 
Discovered by \cite{Rest05b}.
& 

& 
 
& 

SNe~Ia (Rest et al., in preparation).
& 
 
\\
\hline
SNR N103B & 

SN~Ia
& 
 
Discovered by \cite{Rest05b}.
& 

& 
 
& 

SNe~Ia (Rest et al., in preparation).
& 
 
\\
\hline
Tycho & 

SN~Ia
& 
 
Discovered by \cite{Rest07,Rest08b}.
&

&

& 

Classified as normal SN~Ia by \cite{Krause08b}.
&

\\
\hline
Cas~A & 

SN IIb
& 
 
Discovered by \cite{Rest07,Rest08b,Krause08a}.
& 
3D-structure of the absorbing dust filaments using IR echoes by \cite{Kim08}.

&

& 

Classified as SN~IIb by \cite{Krause08a}.
& 
 
SN asymmetry detected by \cite{Rest11_casaspec}, in agreement with
outflow observed in its SNR.
\\
\hline
\end{tabulary}
\caption{Overview of light echoes in the local Universe
\label{table:overview}}
\end{flushright}
\end{sidewaystable}

\clearpage

\end{document}